\documentclass[conference]{IEEEtran}
\IEEEoverridecommandlockouts
\usepackage{cite}
\usepackage{amsmath,amssymb,amsfonts}
\usepackage{algorithmic}
\usepackage{graphicx}
\usepackage{textcomp}
\usepackage{booktabs}
\usepackage{tabularx}
\usepackage{multirow}
\usepackage{makecell}
\usepackage{subfig}

\usepackage{array} 
\usepackage{xcolor}
\def\BibTeX{{\rm B\kern-.05em{\sc i\kern-.025em b}\kern-.08em
    T\kern-.1667em\lower.7ex\hbox{E}\kern-.125emX}}

\newcommand{\ourwork}{ME-ViT}

\begin{document}



\title{ME-ViT: A Single-Load Memory-Efficient FPGA Accelerator for Vision Transformers}

\author{
\IEEEauthorblockN{Kyle Marino}
\IEEEauthorblockA{\textit{University of Southern California} \\
Los Angeles, USA\\
kmarino@usc.edu}
\and
\IEEEauthorblockN{Pengmiao Zhang}
\IEEEauthorblockA{\textit{University of Southern California} \\
Los Angeles, USA\\
pengmiao@usc.edu}
\and
\IEEEauthorblockN{Viktor K. Prasanna}
\IEEEauthorblockA{\textit{University of Southern California} \\
Los Angeles, USA\\
prasanna@usc.edu}
}

\maketitle

\begin{abstract}

Vision Transformers (ViTs) have emerged as a state-of-the-art solution for object classification tasks. However, their computational demands and high parameter count make them unsuitable for real-time inference, prompting the need for efficient hardware implementations. Existing hardware accelerators for ViTs suffer from frequent off-chip memory access, restricting the achievable throughput by memory bandwidth. In devices with a high compute-to-communication ratio (e.g., edge FPGAs with limited bandwidth), off-chip memory access imposes a severe bottleneck on overall throughput. This work proposes ME-ViT, a novel \underline{M}emory \underline{E}fficient FPGA accelerator for \underline{ViT} inference that minimizes memory traffic. We propose a \textit{single-load policy} in designing ME-ViT: model parameters are only loaded once, intermediate results are stored on-chip, and all operations are implemented in a single processing element. To achieve this goal, we design a memory-efficient processing element (ME-PE), which processes multiple key operations of ViT inference on the same architecture through the reuse of \textit{multi-purpose buffers}. We also integrate the Softmax and LayerNorm functions into the ME-PE, minimizing stalls between matrix multiplications. We evaluate ME-ViT on systolic array sizes of 32 and 16, achieving up to a 9.22$\times$ and 17.89$\times$ overall improvement in memory bandwidth, and a 2.16$\times$ improvement in throughput per DSP for both designs over state-of-the-art ViT accelerators on FPGA. ME-ViT achieves a power efficiency improvement of up to 4.00$\times$ (1.03$\times$) over a GPU (FPGA) baseline. ME-ViT enables up to 5 ME-PE instantiations on a Xilinx Alveo U200, achieving a 5.10$\times$ improvement in throughput over the state-of-the art FPGA baseline, and a 5.85$\times$ (1.51$\times$) improvement in power efficiency over the GPU (FPGA) baseline.

\end{abstract}


\section{Introduction}

The self-attention based model of the Transformer\cite{vaswani2017attention} has led to significant advancements in machine learning, impacting a diverse range of applications\cite{arnab2021vivit, chang2022maskgit, hudson2022generative,zhang2022fine,9926307}. 
Originally gaining prominence due to its remarkable success in natural language processing\cite{devlin2018bert}, the Transformer has been adapted to the domain of computer vision via Vision Transformers (ViTs) \cite{dosovitskiy2021image}, achieving superior performance over convolutional networks \cite{liu2021swin, touvron2021training}. Despite the substantial achievements of ViTs, their implementation on real-time image data poses considerable computational and memory challenges due to the immense parameter counts associated with these models \cite{li2022efficientformer}.

Significant effort has been put into accelerating the inference of ViTs, including model size reduction~\cite{mehta2022mobilevit,wu2022tinyvit}, weight quantization~\cite{bhandare2019efficient}, and algorithm optimization~\cite{you2022vitcod}. However, these methods do not directly address the main performance bottleneck of ViT inference on modern hardware: \textbf{memory bandwidth}. Computing capabilities of hardware such as GPUs and TPUs have outpaced memory bandwidth improvements, resulting in a poor Compute-to-Communication (C2C) ratio and limited model performance\cite{ivanov2021data, khan2022hvac}. Various works \cite{dao2022flashattention, tabani2021improving, 8547541} have focused on algorithmic approaches to reducing memory bandwidth, but are still constrained from the overall architectural limitations imposed by the GPU. 


\begin{figure}[t]
    \centering
    \includegraphics[width=0.97\linewidth]{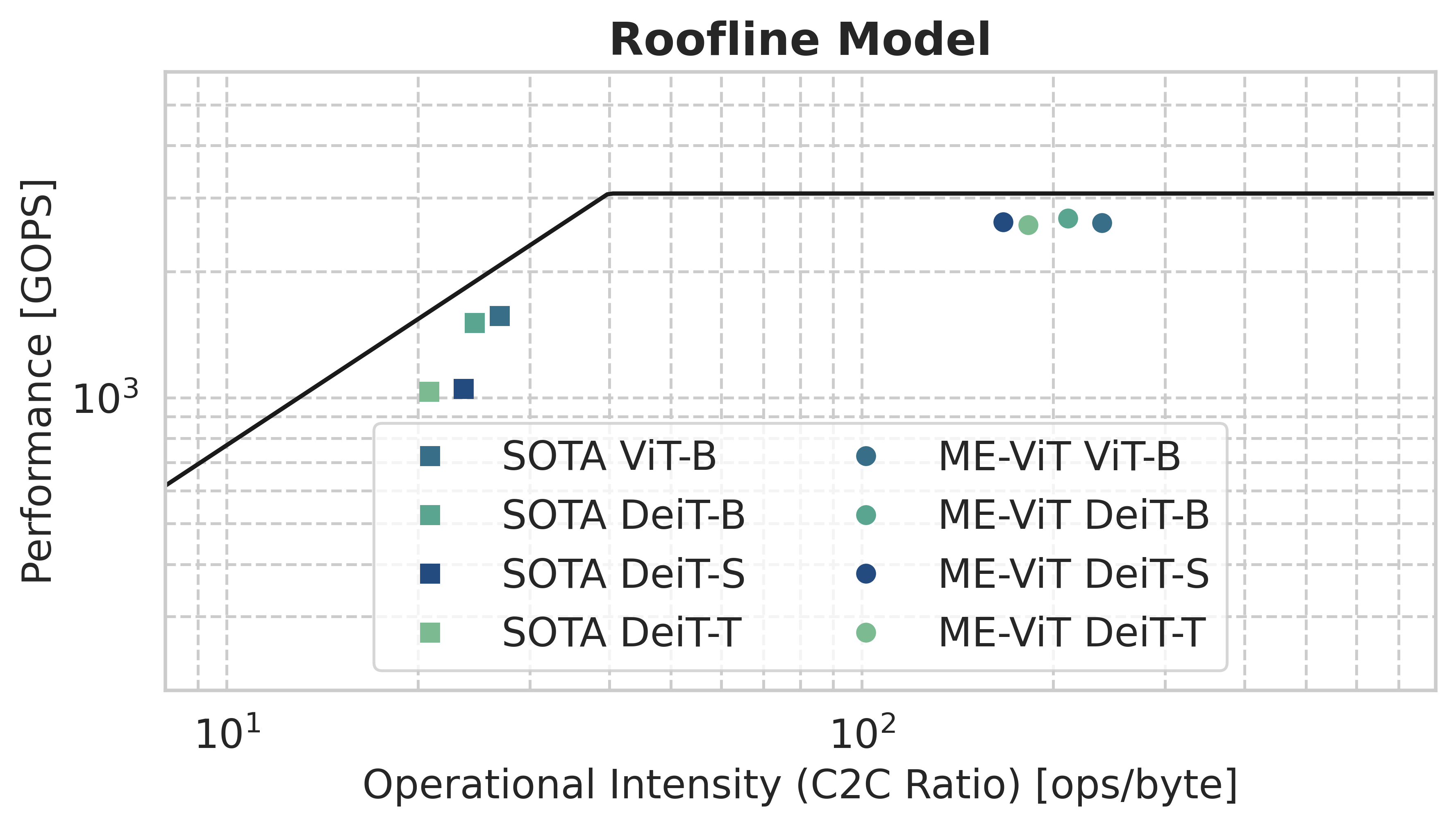}
    \caption{Roofline model of state-of-the-art (SOTA) architectures and ME-ViT for various models. Vertical axis is in log scale. ME-ViT optimizes memory bandwidth, enabling nearly peak performance in GOPS (Giga Operations per Second). SOTA implementations are bottlenecked by memory bandwidth. Four ViT variants are shown: ViT-Base model from \cite{dosovitskiy2021image} (ViT-B), and three models from \cite{touvron2021training} (DeiT-B, DeiT-S, and DeiT-T).}
    \label{fig:roofline}
\end{figure}

Field Programmable Gate Arrays (FPGAs) provide a good platform for ViT acceleration due to the high computational parallelism and custom architectures that can be designed~\cite{hu2021vistop,li2022autovitacc,sun2022vaqf}. However, like GPUs, the high memory bandwidth needs of Transformers significantly limits their implementation on FPGAs. As shown in Figure~\ref{fig:roofline}, ViT and DeiT (a commly-used variant of ViT)~\cite{touvron2021training}  models are severely bottlenecked by memory bandwidth without specialized optimization. 
The shortcomings of modern computing devices for memory-bound computing tasks such as ViT inference prompt the need for efficient and model-specific architectures suited for these tasks.
Custom FPGA architectures can meet the computational and memory demands of ViT inference more effectively than the general-purpose architecture of a GPU, while also consuming less power.

In this work, we aim to minimize the memory bandwidth for ViT inference on an FPGA. The development of such an optimization presents multiple unique challenges. 
\textbf{1) Avoiding write-backs and reloads in block matrix multiplications.} 
The large matrix multiplications present in ViTs require a block matrix multiplication (BMM) approach, which divides the large matrices into smaller blocks to meet the limited resources of an FPGA. 
This results in constant block write-backs and reads to off-chip memory, leading to high usage of memory bandwidth. To minimize the memory traffic, buffering all data on the FPGA is ideal. 
However, excessively large buffers can hinder effective utilization of available DSPs\cite{7459526}. This is because too much buffering per systolic array will exhaust BRAM before all DSPs can be utilized.
Thus, it is crucial to strategically reuse on-chip buffers, achieving a balance between minimal memory traffic and optimal FPGA DSP utilization.
\textbf{2) Buffering intermediate results for the residual connections in ViT.} The residual connections add the values from the previous layer to the computed result of the current layer, which necessitates either the buffering or loading of a previous layer \cite{9925700}. Buffering a layer uses more FPGA resources but reduces the memory traffic that comes with layer loading. Therefore, to minimize memory traffic and to efficiently handle these residual connections, the design of reusable buffers becomes a crucial aspect of the overall system architecture.
\textbf{3) Reducing communications between FPGA accelerator and the host CPU.} Often, Softmax\ and LayerNorm operations in all layers of a ViT are performed on the host CPU, with only the matrix multiplications being offloaded to the accelerator. This is due to the computational insignificance of Softmax and LayerNorm compared to matrix multiplication \cite{sun2022vaqf} \cite{li2022autovitacc}. However, this causes frequent write-backs multiple times per layer, as well as a significant delay for the round trip to the CPU. Constant data transfer becomes a significant performance bottleneck, especially for larger FPGAs with higher processing capabilities. 

We propose~\ourwork, a novel Memory-Efficient ViT accelerator on FPGA that addresses the above challenges. As shown in Figure~\ref{fig:roofline},~\ourwork~optimizes memory bandwidth and enables nearly peak performance for ViTs.~\ourwork~is developed through two key optimizations. 1) A \textbf{Single-load policy} for model parameters loaded from off-chip memory. 2) \textbf{Multi-purpose buffers} for three key operations in ViT.

We propose a \textit{single-load policy} as the key approach for minimizing memory traffic for ViT accelerators. It consists of three objectives. 
First, parameters loaded to the FPGA are only loaded once from off-chip memory (e.g., DRAM). If the value needs to be reused it is strategically buffered in the FPGA's on-chip memory (e.g., BRAM). Second, intermediate data is not written back to off-chip memory between layers. This ensures that the absolute minimum number of memory transfers are used per inference. 
Third, all operations—including BMM, LayerNorm, Softmax, and activations—are performed within a single processing element to eliminate external data traffic. The single-load policy is different than a block  matrix multiplication approach which still requires reloading of the same weights between different blocks.

\textit{Multi-purpose buffers} are designed to achieve the goals of the single-load policy while addressing its requirement of large BRAM allocations.
To minimize resource usage, we design ME-PE, a single \underline{M}emory \underline{E}fficient \underline{P}rocessing \underline{E}lement that executes the three key operations in ViT inference: Linear Projection (LP), Multi-headed Self-Attention (MSA), and Multi-Layer Perceptron (MLP). By designing a custom PE that conforms specifically to the MSA operations, we can strategically order the computations to obtain a minimal resource utilization. The minimal architecture needed for MSA can be repurposed for MLP calculation without requiring additional BRAM resources. BRAM is efficiently packed and repurposed for different stages of calculation. As a result, we are able to design a flexible PE that performs all operations for ViT inference while using minimal BRAM to ensure model parameters are only loaded once and all intermediate data write-back is avoided. Meanwhile, the reuse of resources within our design enables the implementation of multiple ME-PEs in an FPGA, further improving the throughput of ME-ViT.

Our main contributions are summarized as follows:

\begin{itemize}

\item We propose ME-ViT, a novel memory-efficient Vision Transformer accelerator on FPGA, which optimizes memory bandwidth and achieves nearly peak performance in operations per second.

\item We propose a single-load policy as the core principle for minimizing memory access by only loading data once from DRAM, buffering intermediate results, and implementing all operations in a single PE. 

\item We design an ME-PE, a memory-efficient processing element with reusable multi-purpose buffers. This novel PE enables three key operations of ViT inference to be processed on the same architecture, minimizing resource usage and retaining intermediate data between operations.

\item We integrate LayerNorm and Pseudo-Softmax (a hardware-optimized Softmax) in the ME-PE to avoid off-chip computation and reduce data traffic. These functions are pipelined with matrix multiplication to reduce computational stalls.

\item We evaluate ME-ViT on a Xilinx Alveo U200. Using systolic array sizes of 32 and 16, ME-ViT achieves up to a 9.22$\times$ and 17.89$\times$ overall improvement in memory bandwidth, and up to a 2.16$\times$ improvement in throughput per DSP over state-of-the-art ViT accelerators on FPGA. ME-ViT achieves up to 4.00$\times$ (1.03$\times$) power efficiency over the GPU (FPGA) baseline. The ME-ViT with 5 ME-PEs implemented on board achieves a 5.1$\times$ improvement in throughput over the state-of-the-art FPGA baseline, and a 5.85$\times$ (1.51$\times$) improvement in power efficiency over the GPU (FPGA) baseline.


\end{itemize} 

To the best of our knowledge, ME-ViT is the first Vision Transformer architecture that optimizes memory traffic and strictly enforces a minimal memory access policy. 

\section{Background}
\label{sec:background}
\subsection{Vision Transformer}
\begin{figure}[t]
  \centering
  \includegraphics[width=\linewidth]{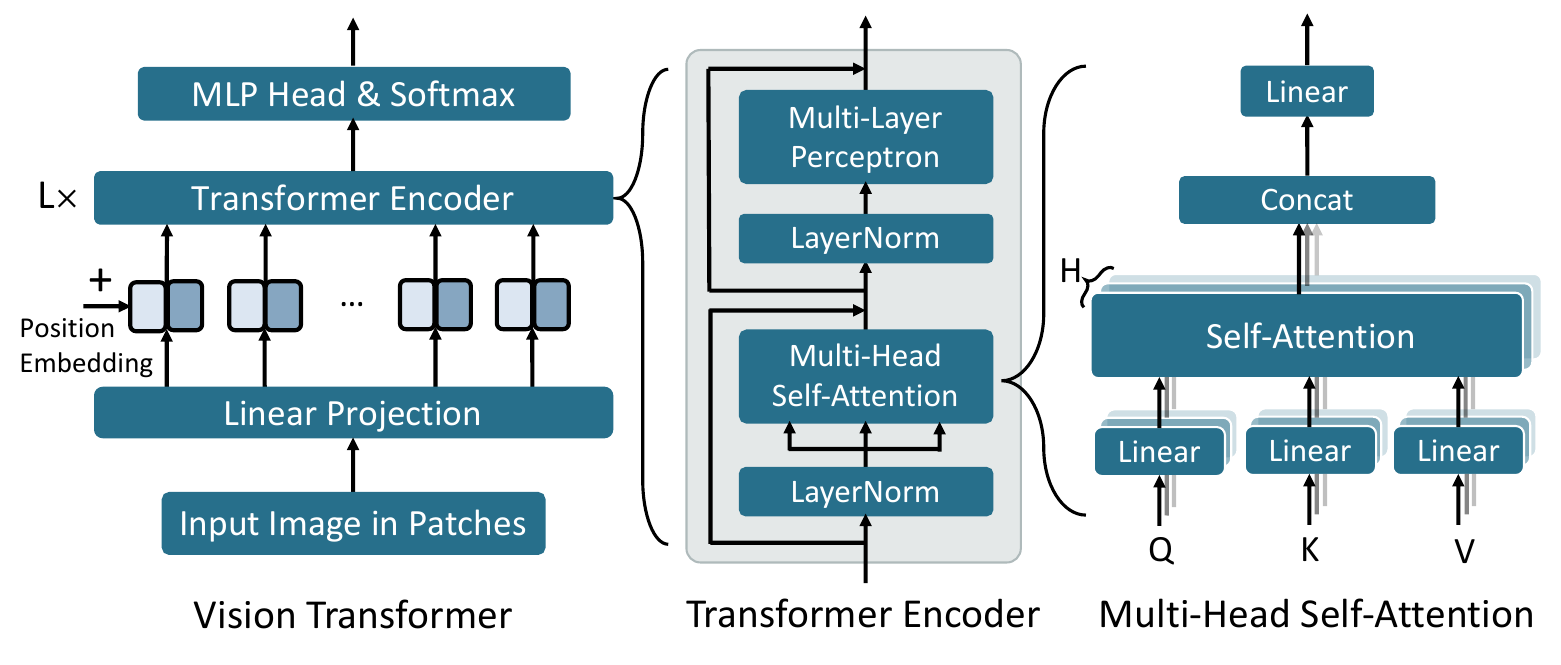}
  \caption{Vision Transformer architecture.}
\label{fig:vit}
\end{figure}

Vision Transformer (ViT) is a state-of-the-art deep learning architecture that utilizes the Transformer model for computer vision tasks. By leveraging self-attention mechanisms, ViTs have achieved significant advancements in image classification and object detection, revolutionizing the field of visual recognition.
The ViT architecture is shown in Figure~\ref{fig:vit}. The input image is broken up into patches and fed into the Transformer Encoder as a sequence. The Transformer Encoder is mainly composed of a multi-headed self-attention block (MSA), a multi-layer perceptron block (MLP), and layer normalization blocks (LN).

\noindent\textbf{Multi-Headed Self-Attention.}
Self-attention takes the embedding of items as input, converts them to three matrices through linear projection, then feeds them into a scaled dot-product attention. The self-attention function is defined as:
\begin{equation}
\label{eq:att1}
\operatorname{Attention}(Q, K, V)=\operatorname{softmax}\left(\frac{Q K^{\top}}{\sqrt{D_{k}}}\right) V
\end{equation}
where $\mathrm{Q}$ is queries, $\mathrm{K}$ is keys, $\mathrm{V}$ is values, $\mathrm{D}$ is the model dimension, and $\mathrm{D_{k}}$ is the dimension of $\mathrm{K}$. Considering one self-attention operation as one "head," Multi-headed Self-Attention (MSA) operation is shown as:
\begin{equation}
\begin{aligned}
\operatorname{MSA}(Q, K, V) &=\operatorname{Concat}\left(\operatorname{head}_{1}, \ldots, \text {head}_{\mathrm{h}}\right) W^{O} \\
\text{head}_{\mathrm{i}} &=\text {Attention}\left(\mathrm{Q W_{i}^{Q}, K W_{i}^{K}, V W_{i}^{V}}\right)
\end{aligned}
\end{equation}
where the projection matrices $\mathrm{W_{i}^{Q}, W_{i}^{K}, W_{i}^{V} \in \mathbb{R}^{D \times D_{h}}}$, $\mathrm{h}$ is the total number of heads, $\mathrm{i}$ is the index of heads, and $\mathrm{D_h=D/h}$ is the dimension of each head.

\noindent\textbf{Multi-Layer Perceptron.}
The MLP block consists of two linear layers with an activation function:

\begin{equation}
    \operatorname{MLP}(x) = \operatorname{GeLU}(xW^H + B^H)W^O + B^O
\end{equation}
where $\mathrm{W^H}$ is the hidden layer weights, $\mathrm{B^H}$ is the hidden layer bias, $\mathrm{W^O}$ is the output layer weights, and $\mathrm{B^O}$ is the output layer bias, $\operatorname{GeLU}$ is an activation function\cite{hendrycks2016gaussian}.


\noindent\textbf{Transformer Encoder.}
For the input image $\mathrm{\mathbf{x} \in \mathbb{R}^{H \times W \times C}}$, the formal expression of transformer encoder layers are shown in Equation~\ref{eq:vit}.
\begin{equation}
\begin{aligned}
\label{eq:vit}
\mathbf{z}_0 & =\left[\mathbf{x}_{\text {class}} ; \mathbf{x}_p^1 \mathbf{E} ; \mathbf{x}_p^2 \mathbf{E} ; \cdots ; \mathbf{x}_p^N \mathbf{E}\right]+\mathbf{E}_{p o s}\\
\mathbf{z}_{\ell}^{\prime} & =\operatorname{MSA}\left(\operatorname{LN}\left(\mathbf{z}_{\ell-1}\right)\right)+\mathbf{z}_{\ell-1}\\
\mathbf{z}_{\ell} & =\operatorname{MLP}\left(\operatorname{LN}\left(\mathbf{z}_{\ell}^{\prime}\right)\right)+\mathbf{z}_{\ell}^{\prime}\\
\mathbf{y} & =\operatorname{LN}\left(\mathbf{z}_L\right) 
\end{aligned}
\end{equation}
where $\mathrm{\mathbf{x}_{class}}$ is a class token, $\mathrm{\mathbf{x}_p \in \mathbb{R}^{N \times (P^2 \times C)}}$ is the image segmented to $\mathrm{N}$ patches, $\mathrm{\mathbf{E} \in \mathbb{R}^{\left(P^2 \cdot C\right) \times D}}$ is the input embedding, $\mathrm{\mathbf{E}_{p o s} \in \mathbb{R}^{(N+1) \times D}}$ is the position embedding, $\mathrm{\ell=1 \ldots L}$ is the index of layers.

In this work, we design a ViT accelerator by implementing the above functions on FPGA and optimizing the memory accesses in model inference.


\begin{figure}[t]
  \centering
  \includegraphics[width=0.8\linewidth]{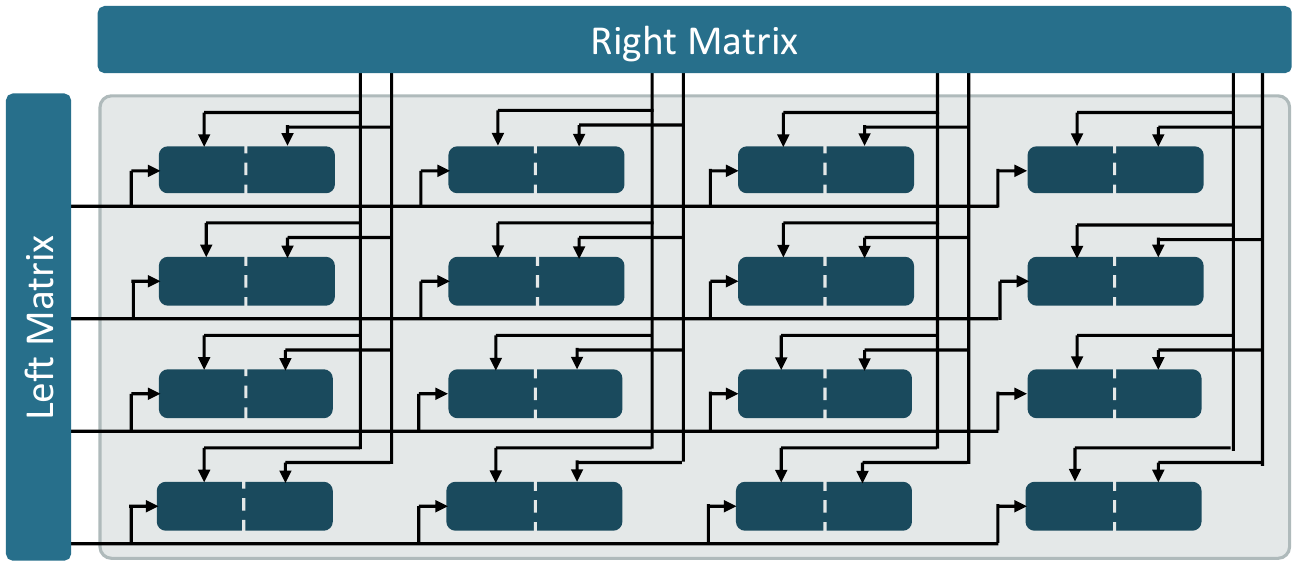}
  \caption{Systolic array architecture of size 4$\times$4. Dark gray boxes indicate DSPs, each split by a dashed line to illustrate DSP packing.}
\label{fig:sys-array}
\end{figure}

\subsection{ViT Accelerators on FPGA}

Field Programmable Gate Arrays (FPGAs) have been extensively used for accelerating machine learning tasks \cite{nair2023fpga}\cite{iskandar2022near}. FPGAs consist of a programmable interconnect of logic gates and on-chip memories (BRAMs, URAMs, LUTRAMs), allowing custom hardware architectures to be designed. This flexibility enables hardware optimizations to meet the specific requirements of various machine learning tasks. 


There have been several proposed architectures for accelerating ViT and, more generally, Transformer inference on FPGAs \cite{lu2020hardware, sun2022vaqf, 9925700, nag2023vita}. These architectures typically quantize weights and activations to 8 bits to reduce model size and computation requirements \cite{li2022autovitacc, sun2022vaqf}. Some architectures compute Softmax and LayerNorm on the FPGA \cite{nag2023vita, 9925700}, while others perform this on the host CPU \cite{sun2022vaqf}. Computing these functions on the CPU reduces the complexity of the FPGA design, but increases the memory overhead. Wang et al. \cite{9925700} propose ViA, a ViT accelerator which performs the full calculation per layer on an FPGA but requires write-backs between layers and does not implement the full-size ViT models. Nag et al. \cite{nag2023vita} target an edge FPGA device and also incur frequent memory reads and write-backs between layers. Sun et al. \cite{sun2022vaqf} propose an accelerator design that requires the CPU to perform the Softmax and LayerNorm functions and also necessitates loading and unloading for each matrix multiplication. Lu et al. \cite{lu2020hardware} propose an accelerator for the MSA and MLP layers of a Transformer with on-chip computation of Softmax and LayerNorm, but require memory access between matrix multiplications.

All of the above model architectures write back intermediate layer calculations and frequently load weights for matrix multiplication, which incur a large memory overhead \cite{lu2020hardware, 9925700, nag2023vita, sun2022vaqf}. As these architectures scale to larger sizes, they become limited by memory bandwidth constraints. ME-ViT aims to address these memory constraints by eliminating intermediate write backs between layers and only loading weights once into the design, ensuring the minimal amount of data is transferred. This policy enables our design to easily scale without being constrained by memory limitations.

\subsection{Matrix Multiplication on FPGA}

\subsubsection{DSP Packing}
DSP packing is utilized to perform two simultaneous multiplies per DSP as described in \cite{dsp-pack}. Each DSP contains an 18-bit $\mathrm{\times}$ 27-bit multiplier, which can be used to perform $\mathrm{A \times B}$ and $\mathrm{A \times C}$ simultaneously by assigning $\mathrm{A}$ to the 18-bit operand, and $\mathrm{(B<<18 + C)}$ to the 27-bit operand. The resulting multiply leaves the two distinct 16-bit products separated in the 45-bit DSP output. A systolic array of size $\mathrm{P_{SYS} \times P_{SYS}}$ shown in Figure~\ref{fig:sys-array} is used to efficiently multiply matrices. With DSP packing, the systolic array can perform $\mathrm{P_{SYS} \times (2 \cdot P_{SYS})}$ multiplies. Datapaths in Figures~\ref{fig:lp-arch},~\ref{fig:msa-arch}, and~\ref{fig:mlp-arch} with double-arrows indicate the double-wide data paths, which are necessary for accommodating the combined bit width of the two separate products.


\begin{figure}[t]
\centering
\subfloat[MLP Matrix Multiplication]
{\includegraphics[width=1\linewidth]{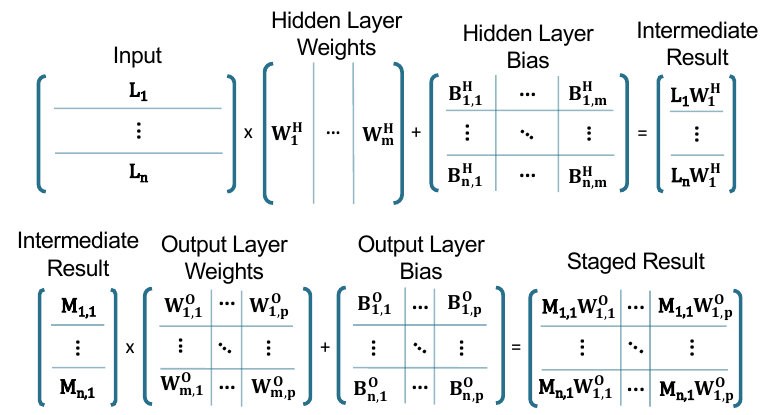}}
\newline
\subfloat[Partial Sum Method]
{\includegraphics[width=0.9\linewidth]{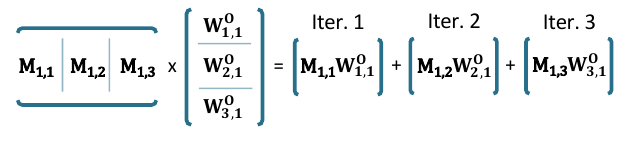}}
\caption{\centering MLP matrix multiplication and partial sum method.}
\label{fig:mlp-method}
\end{figure}


\subsubsection {Matrix Multiplication for MLP}
The MLP calculation involves multiplication with weight matrices $\mathrm{W^H}$ and $\mathrm{W^O}$, which are too large to be buffered or computed fully within the PE. Matrix multiplication is split into row blocks and column blocks, where $\mathrm{A_i}$ refers to the $\mathrm{i}$th row or column block of matrix $\mathrm{A}$. A sub-block $\mathrm{B_{i,j}}$ refers to the matrix block at the $\mathrm{i}$th row block and $\mathrm{j}$th column block of matrix $\mathrm{B}$.

To perform the calculation without reloading weights, we implement the MLP method from  \cite{nag2023vita} shown in Figure~\ref{fig:mlp-method}. A single sub-block of the intermediate result is calculated, which then is passed point-wise through an activation function. The result $\mathrm{M_{i,j}}$ is multiplied by each sub-block in $\mathrm{W^O}$. The result of each block multiplication is added to the previous value in the staged result. At the start of the cycle, the staged result contains the output layer bias, which handles bias addition implicitly. This performs the partial sum method shown in Figure \ref{fig:mlp-method}b, which allows the $\mathrm{M}$ row block and $\mathrm{W^O}$ column block product to be broken up into separate iterations. For each iteration in Figure \ref{fig:mlp-method}a, a column block from $\mathrm{W^H}$ and a row block from $\mathrm{W^O}$ is loaded. This approach allows the weights in the MLP layer to be fully used from a single load without requiring larger buffer sizes.

\section{Approach}
\label{sec:approach}
{\color{black}
We introduce ME-ViT, a novel memory-efficient FPGA accelerator for Vision Transformers. ME-ViT minimizes memory traffic, which is accomplished through a versatile architecture of a memory-efficient processing element (ME-PE, Section~\ref{sec:me-pe}), the scheduling of ME-PE modes for the key operations in a Transformer Encoder (LP mode in Section~\ref{sec:mode-lp}, MSA mode in Section~\ref{sec:mode-msa}, and MLP mode in Section~\ref{sec:mlp-mode}), and the integration of LayerNorm and Softmax operations (Section~\ref{sec:layernorm_softmax}). Due to the high reuse of resources, we design an ME-ViT architecture with multiple ME-PEs working in parallel (Section~\ref{sec:multi-pe}) that highly improves the throughput of ViT inference.
}







\begin{figure}[t]
    \centering
    \includegraphics[width=0.87\linewidth]{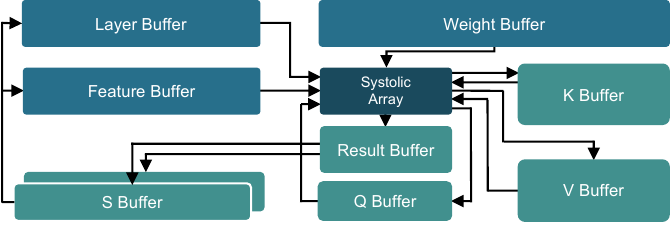}
    \caption{\centering Overview of the ME-PE architecture.}
    \label{fig:me-pe-overview}
\end{figure}

\begin{figure}[t]
    \centering
    \includegraphics[width=0.95\linewidth]{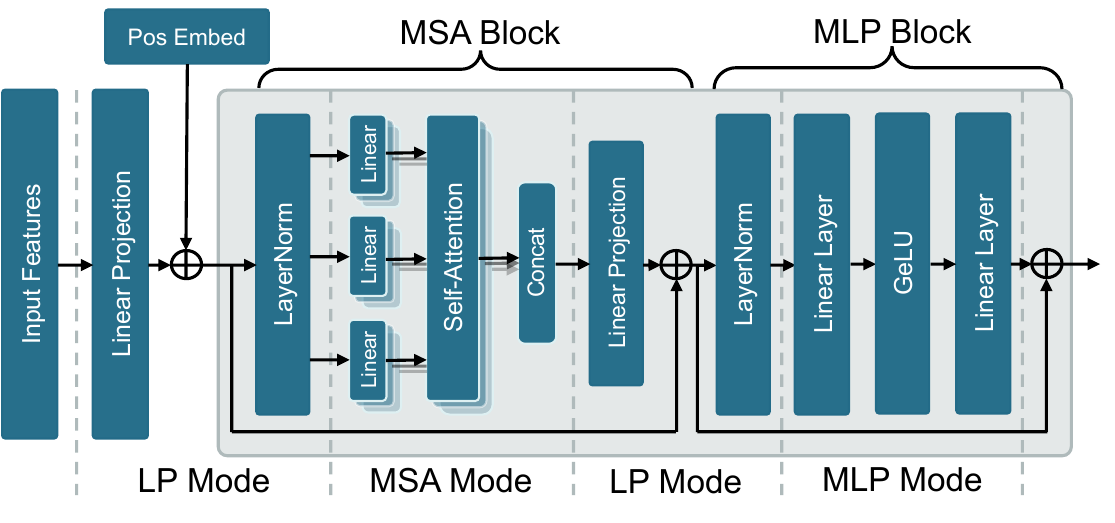}
    \caption{Scheduling of ME-PE modes across Transformer encoder for ME-ViT inference.}
    \label{fig:mode-sched}
\end{figure}

\begin{figure*}[ht]
    \centering
    \includegraphics[width=1\linewidth]{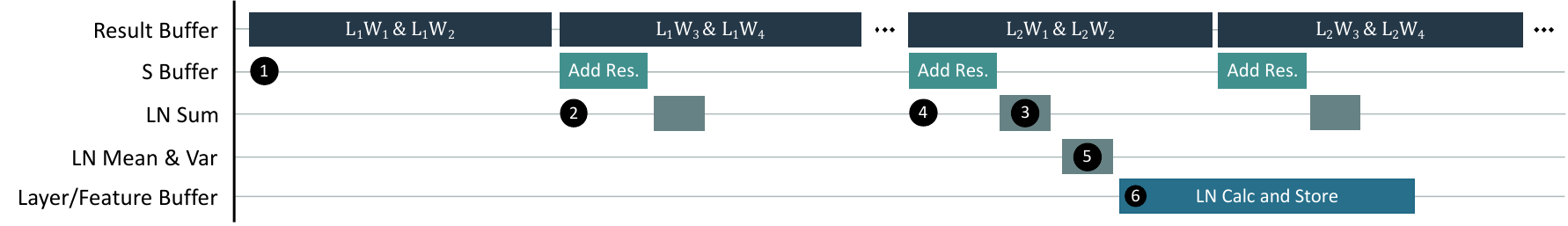 }
    \caption{Scheduling for Linear Projection Mode.}
    \label{fig:lp-sch}
\end{figure*}

\begin{figure}[t]
    \centering
    \includegraphics[width=0.9\linewidth]{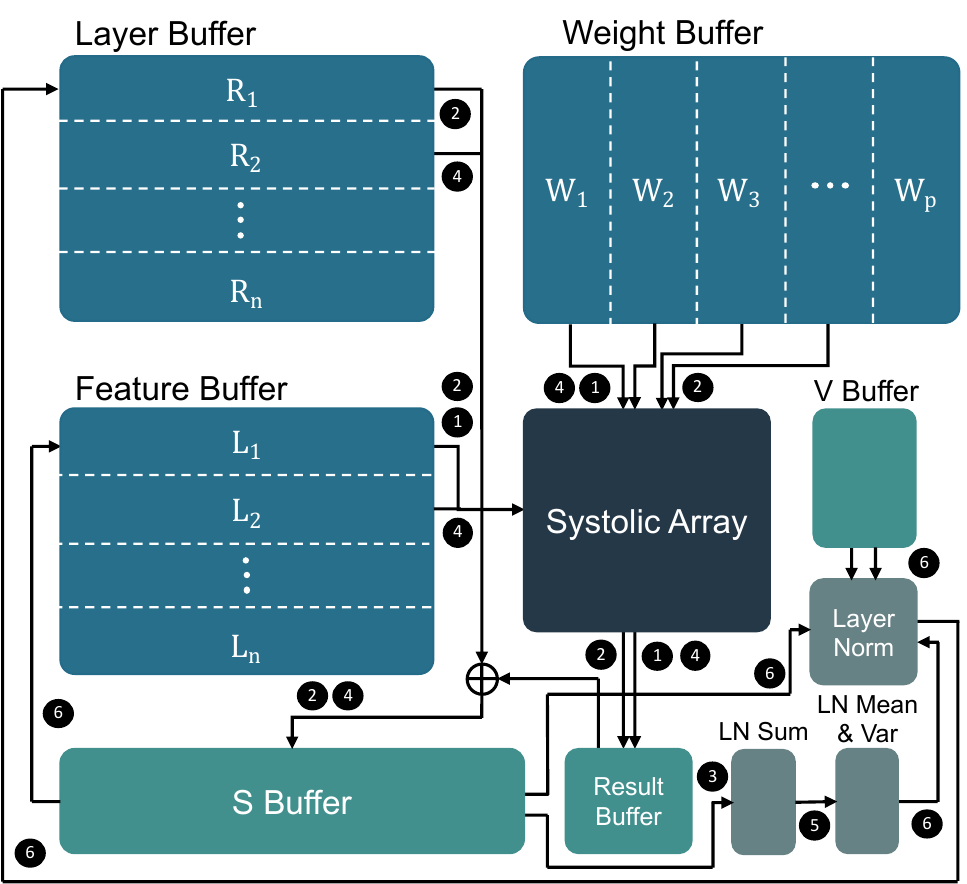 }
    \caption{Linear Projection process on ME-VIT. }
    \label{fig:lp-arch}
\end{figure}

\subsection{Architecture of Memory-Efficient Processing Element}
\label{sec:me-pe}

The Memory-Efficient Processing Element (ME-PE) is designed such that distinct input parameters such as weights and biases are only loaded once into the design from off-chip memory (e.g., DRAM). In addition, there is no unloading of any intermediate values. These two objectives of the ME-PE ensure that the minimal amount of memory traffic is incurred. A final objective of this design is to minimize the number of computation stalls for internal data movement within the PE. The hardware architecture of the ME-PE is shown in Figure~\ref{fig:me-pe-overview}, centered around a systolic array with multi-purpose buffers to manage data around it.

There are 3 modes of operation for the ME-PE shown in Figure \ref{fig:mode-sched}: 1) Linear Projection (LP) Mode; 2) Multi-headed Self-Attention (MSA) Mode; and 3) Multi-Layer Perceptron (MLP) Mode. Each ME-PE handles all modes on the same architecture with only one extra buffer created to handle MLP Mode (see Section~\ref{sec:mlp-mode}). Since each ME-PE performs all calculations needed for ViT inference, individual layer calculations do not need to be written back to memory. In addition, the residual layer connections present in the Transformer architecture are retained within the PE and similarly do not require additional memory access.

\begin{figure*}[ht]
    \centering
    \includegraphics[width=0.95\linewidth]{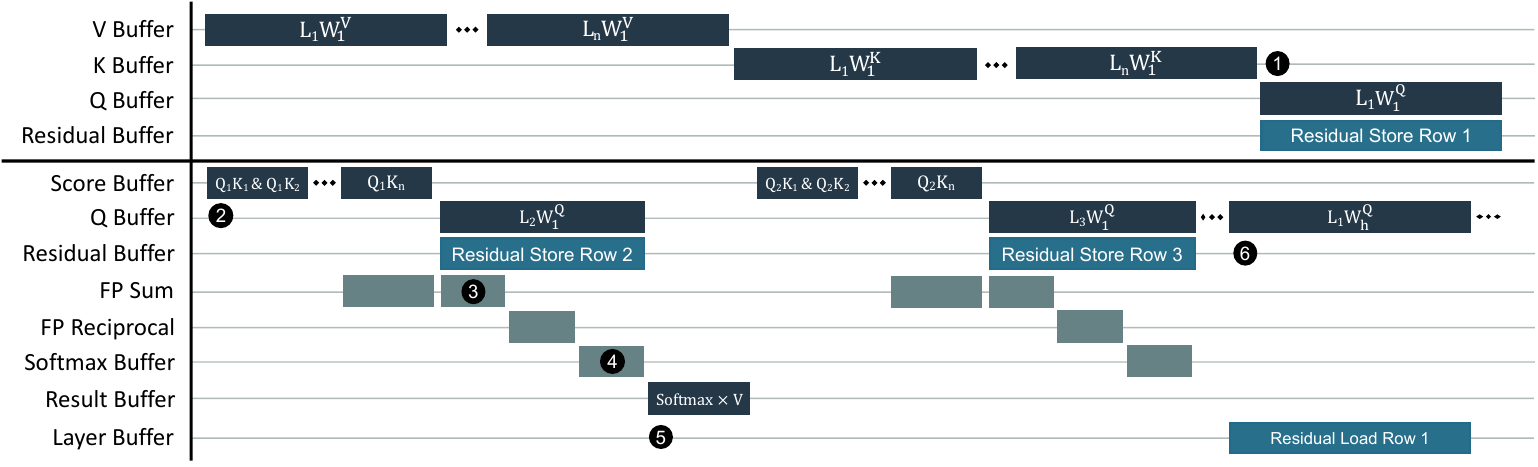}
    \caption{Scheduling for MSA Mode. Top half shows the V and K buffer initialization before the main operation.}
    \label{fig:msa-sch}
\end{figure*}

The ME-PE consists of 3 BRAM buffers and 6 LUTRAM buffers. The Weight Buffer stores a maximum matrix size of $\mathrm{D \times D}$ bytes, and the Feature and Layer Buffers store a maximum matrix of size of $\mathrm{(N+1) \times D}$ bytes. For the base ViT model, $\mathrm{D = 768}$ and $\mathrm{N = 256}$. The Weight Buffer, Layer Buffer, and Feature Buffer are implemented in BRAM and use $\mathrm{160}$, $\mathrm{64}$, and $\mathrm{64}$ 36k BRAMs respectively, for a total of $\mathrm{288}$ BRAMs. While smaller BRAM allocations could store the required data, a multiple of $\mathrm{P_{SYS}}$ must be used to meet the parallel access needs of the systolic array. 

\begin{figure}[t]
    \centering
    \includegraphics[width=0.95\linewidth]{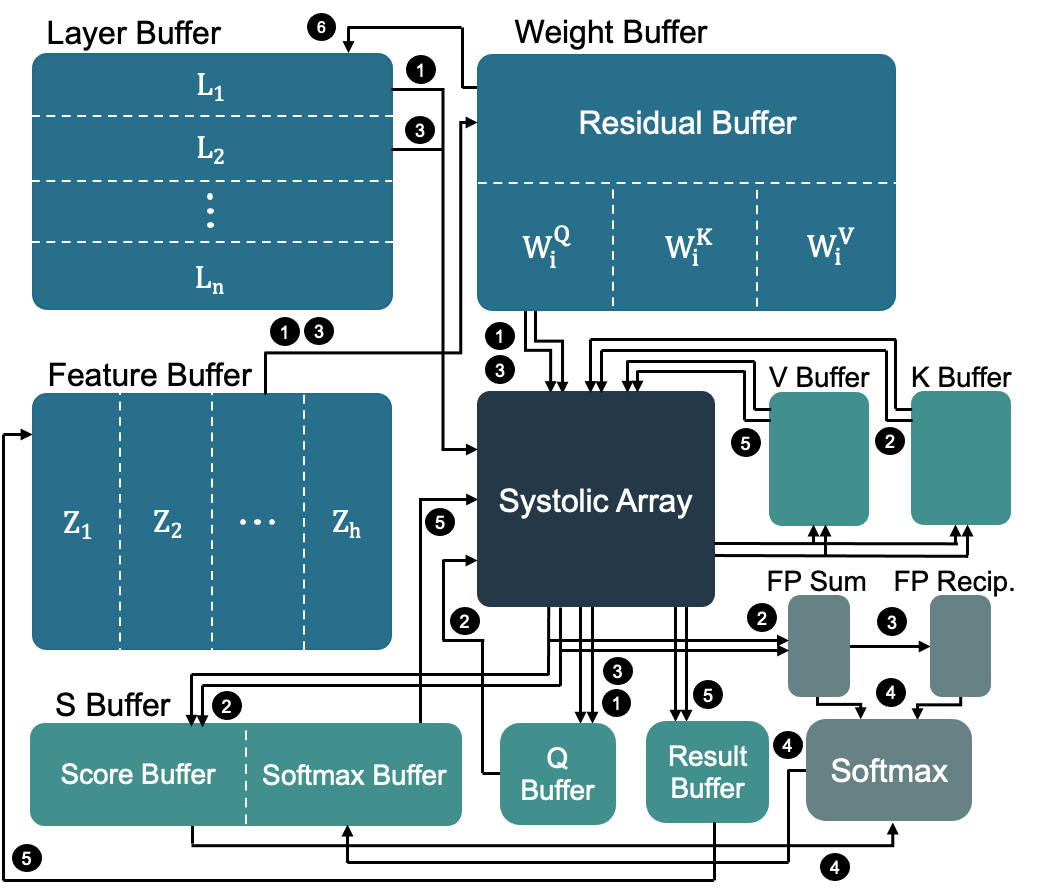}
    \caption{MSA process on the ME-VIT architecture.}
    \label{fig:msa-arch}
\end{figure}

The Q, K, V, Result, and two S Buffers are implemented in LUTRAM due to their small size and parallel data access. The sizes of these buffers are $\mathrm{P_{SYS} \times D_{h}}$, $\mathrm{D \times D_{h}}$, $\mathrm{D \times D_{h}}$, $\mathrm{P_{SYS} \times (2 \cdot P_{SYS})}$, and $\mathrm{P_{SYS} \times D}$ respectively. Most buffers are multi-purpose, and the names suggest the general use case.

\subsection{Linear Projection Mode}
\label{sec:mode-lp}

The Linear Projection (LP) Mode performs BMM to compute Feature Buffer $\mathrm{\times}$ Weight Buffer as shown in Fig~\ref{fig:lp-arch}. This operation is needed for the linear projection of input features and to compute the output linear layer in MSA. In addition to performing matrix multiplication, LP Mode performs the residual connection addition and LayerNorm. The LayerNorm result is stored in the Layer Buffer to be used as an input for the next MSA or MLP block. Since the residual sum is also used as an input to the next block, this sum is stored in the Feature Buffer. 

LP Mode places a minimum requirement for BRAM usage to store both the layer matrix (L) and the weight matrix (W) since these weights are repeatedly accessed during linear projection. A matrix multiplication of the same size occurs in the output linear projection of the MSA block, so this mode is reused at that step.

\begin{figure}[h]
    \centering
    \includegraphics[width=0.82\linewidth]{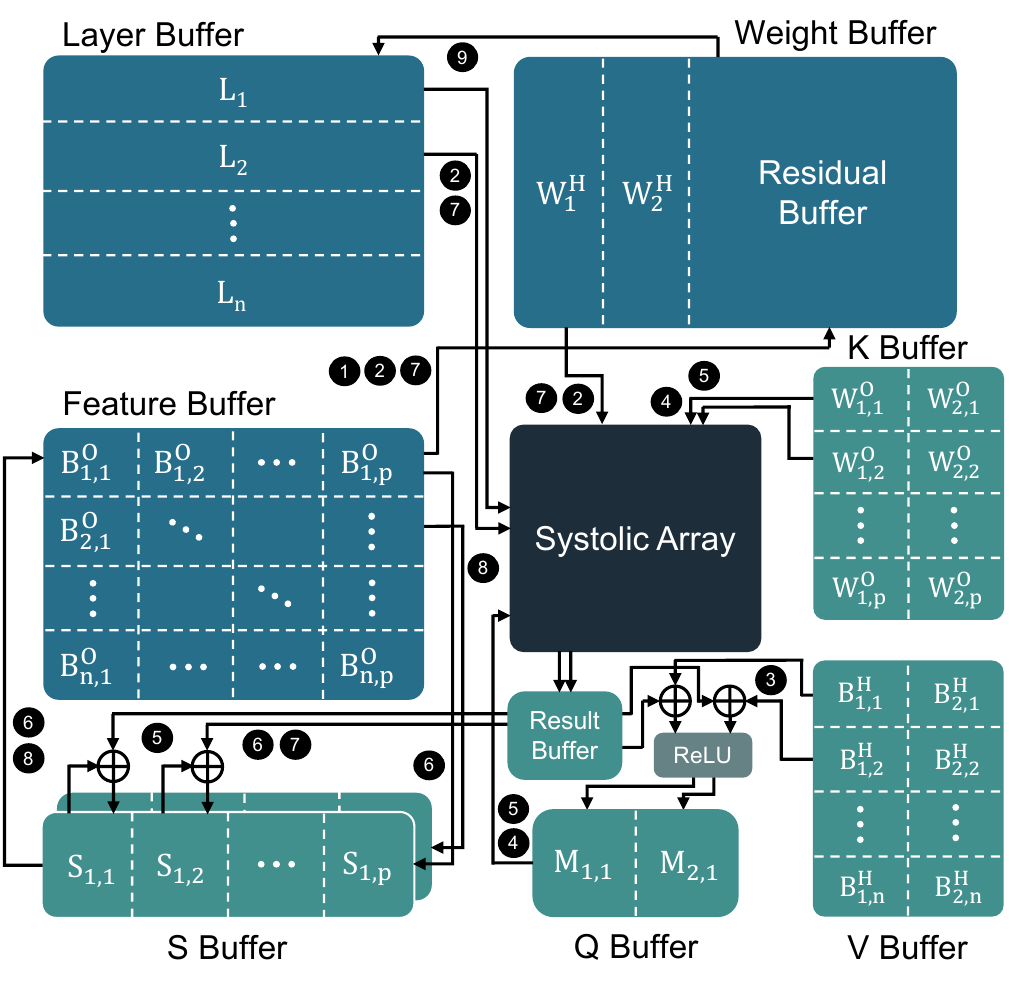}
    \caption{MLP process on the ME-VIT architecture.}
    \label{fig:mlp-arch}
\end{figure}

Figure~\ref{fig:lp-sch} illustrates the task scheduling for this mode. At timestamp 1, the block matrix multiplications $\mathrm{L_1 \times W_1}$ and $\mathrm{L_1 \times W_2}$ are concurrently performed and stored in the Result Buffer. At timestamp 2, the Result Buffer values are added to the residual connection values from the Layer Buffer and stored in the S Buffer. Meanwhile, the next pair of block matrix multiplications are performed. Timestamp 3 calculates the sum and squared sum from the new values in the S Buffer to be later used for LayerNorm. At timestamp 4, the cycle repeats with the next row block from the Feature Buffer. At timestamp 5, the mean and variance are calculated from the sums stored in LN Sum. At timestamp 6, LayerNorm is calculated (see Section~\ref{sec:layernorm}) and the results are stored in the Layer Buffer. The un-normalized values are moved to the Feature Buffer and will be used later for residual connection.

\subsection{Multi-Headed Self-Attention Mode}
\label{sec:mode-msa}
The Multi-Headed Self-Attention (MSA) Mode performs the MSA operation for a single head at a time on the ME-PE architecture shown in Figure~\ref{fig:msa-arch}. At the start of operation, the Layer Buffer contains the LayerNorm result calculated from either the previous LP or MLP operation. The Feature Buffer contains the previous layer result before layer normalization to be used later for the residual connection. Task scheduling of this mode is shown in Figure~\ref{fig:msa-sch}. At the start of operation, the $\mathrm{V_i}$ and $\mathrm{K_i}$ matrices are calculated and stored in the respective buffers. The weight matrices $\mathrm{W^V_i}$, $\mathrm{W^K_i}$, and $\mathrm{W^Q_i}$ are loaded during the first few block matrix multiplication iterations during the $V_i$ calculation. At timestamp 1, the residual data stored in the Feature Buffer is moved to empty space in the Weight Buffer so the head outputs can be stored in the Feature Buffer. Simultaneously, the first block row of $\mathrm{Q}$ for head 1 is calculated by $\mathrm{L_1 \times W^Q_1}$ and stored in the Q Buffer. At timestamp 2, the Q Buffer is multiplied by two block columns from the K Buffer until the full row of the S Buffer is calculated. At timestamp 3, the next row block of Q is calculated, the next residual row is stored, and the FP reciprocal is calculated. At timestamp 4, Softmax is calculated (see Section~\ref{sec:softmax}). The cycle repeats until the last head, where timestamp 5 shows how the residual data is moved to the Layer Buffer as the previously stored data there is no longer needed. This leaves the ME-PE with the output attention matrices stored in the Feature Buffer and residual connection data in the Layer Buffer at the end of the operation.

\subsection{Multi-Layer Perceptron Mode}
\label{sec:mlp-mode}

\begin{figure*}[htp]
    \centering
    \includegraphics[width=\linewidth]{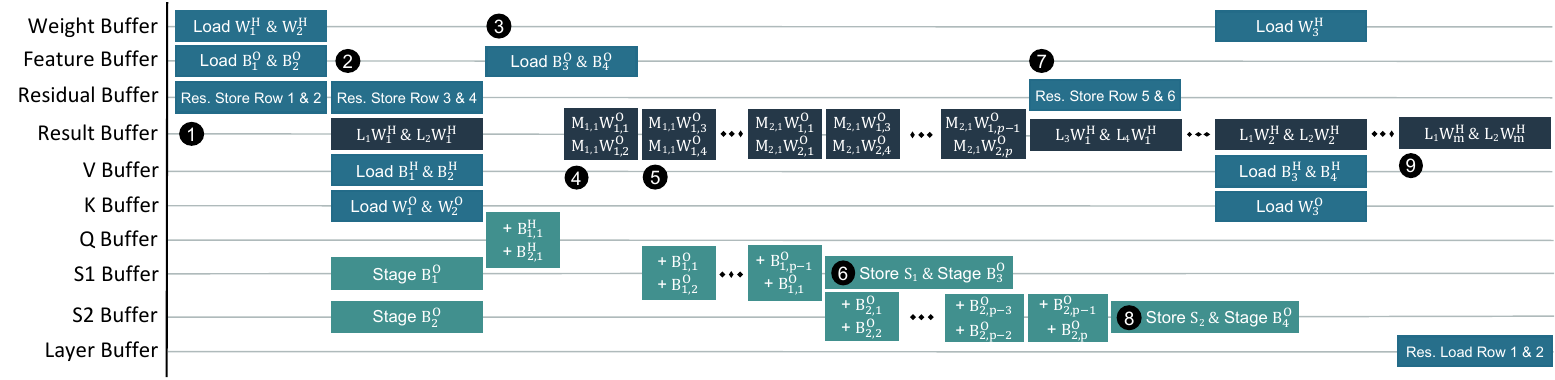}
    \caption{Scheduling for MLP Mode.}
    \label{fig:mlp-sch}
\end{figure*}

\begin{figure}[h]
\centering
\subfloat[LayerNorm Module]
{\includegraphics[width=0.9\linewidth]{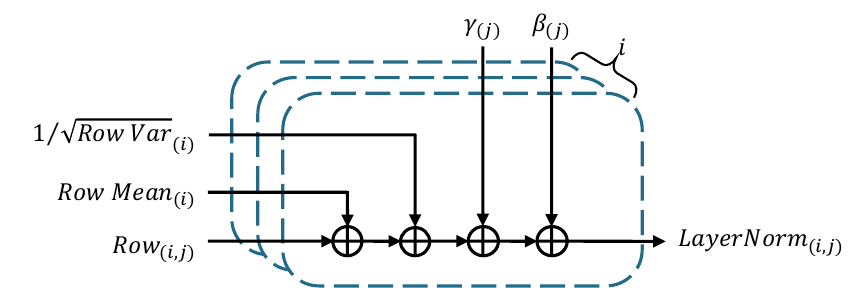}\label{fig:layernorm}}
\newline
\subfloat[Pseudo-Softmax Module]
{\includegraphics[width=0.9\linewidth]{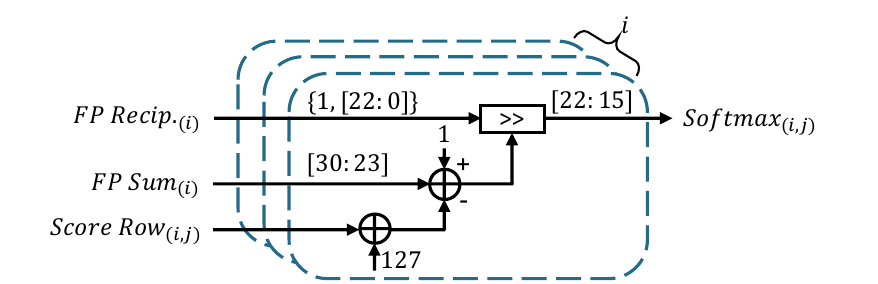}\label{fig:softmax}}
\caption{Design of LayerNorm and Pseudo-Softmax Modules.}
\label{fig:norm_soft}
\end{figure}

The Multi-Layer Perceptron (MLP) Mode performs the MLP operation on the ME-PE shown in Figure~\ref{fig:mlp-arch}. Like with MSA Mode, the Layer Buffer contains the previously calculated LayerNorm, and the Feature Buffer contains the non-normalized values. The operation scheduling in MLP Mode is shown in Figure~\ref{fig:mlp-sch}. At timestamp 1, column blocks $\mathrm{W^H_1}$ and $\mathrm{W^H_2}$ are loaded into the Weight Buffer and row blocks $\mathrm{B^H_1}$ and $\mathrm{B^H_2}$ are loaded into the Feature Buffer. At timestamp 2, the block matrix multiplications $\mathrm{L_1 \times W^H_1}$ and $\mathrm{L_2 \times W^H_1}$ are concurrently performed and stored in the Result Buffer. The column blocks $\mathrm{B^H_1}$ and $\mathrm{B^H_2}$ are loaded to the V Buffer and row blocks $\mathrm{B^H_1}$ and $\mathrm{B^H_2}$ are loaded into the K buffer. The row blocks $\mathrm{B^H_1}$ and $\mathrm{B^H_2}$ are staged by being moved to the S Buffer for future addition. At timestamp 3, the row blocks $\mathrm{B^H_1}$ and $\mathrm{B^H_2}$ are added to the two blocks stored in the Result Buffer. Instead of GeLU, the ReLU activation function is used for hardware simplicity. Various hardware approximations of GeLU exist and would not affect the scheduling or timing of this design. The activation function results are stored in the Q buffer. This timestamp results in a stall of the systolic array, however it is unavoidable since the result is needed immediately for the next multiply. This only incurs a delay of $\mathrm{P_{SYS}}$ clocks which is insignificant. The next pair of output biases are also loaded. At timestamp 4, the first two output layer block multiplications are computed. At timestamp 5, the results computed in timestamp 4 are added to the staged values in the $\mathrm{S_{1}}$ buffer, and the next two output layer blocks are calculated. This process repeats until all sub-blocks involving $\mathrm{M_{1,1}}$ are calculated. At timestamp 6, the $\mathrm{S_{1}}$ buffer containing the staged values is stored back into the Feature Buffer, and multiplication repeats with $\mathrm{M_{2,1}}$. At timestamp 7, the next pair of Layer Buffer row blocks is processed.  This continues until all row blocks in the Layer Buffer have been multiplied by $\mathrm{W^H_1}$. After all layer row blocks have been processed, the cycle repeats with the next $\mathrm{W^H}$ column block. At timestamp 8, $\mathrm{S_{2}}$ is stored and the next row block from the Feature Buffer is loaded to $\mathrm{S_{2}}$. On the last cycle in timestamp 9, the residual data stored in the Weight Buffer is transferred back to the Layer Buffer, replacing the existing values as they are no longer needed. 

Two separate S Buffers are needed to reduce stalls for matrix multiplication. After one S buffer is filled it gets written back to Feature Buffer BRAM. While this takes place, the second S buffer gets written to. Otherwise, matrix multiplication would need to be stalled while the S buffer gets unloaded and reloaded with new values. This adds an additional resource overhead to MLP Mode that is not present for other modes. However, since the S buffer is relatively small and implemented in LUTRAM, this incurs minimal penalty.

\subsection{Integrating LayerNorm and Pseudo-Softmax}
\label{sec:layernorm_softmax}


\subsubsection{LayerNorm Module}

\label{sec:layernorm}

We implement LayerNorm in the ME-PE using a two-pass approach that is performed in parallel with matrix multiplication. LayerNorm is calculated in both the MSA and MLP blocks with:

\begin{equation}
    LayerNorm_{i,j} =  \frac{X_{i,j} - \mu_{i}}{\sqrt{\sigma^2_{i} + \epsilon}} \cdot \gamma_j + \beta_j
\end{equation}
With the mean and variance per $ith$ row of matrix $X$ as:
\begin{equation}
    \mu_{i} = \frac{1}{n}\sum_{j=1}^{n} X_{i,j}
\end{equation}
\begin{equation}
    \sigma^2_i = \frac{1}{n}\sum_{j=1}^{n}(X_{i,j} - \mu_{i})^2\\
\end{equation}
The mean and variance can be calculated in parallel using the variance form:
\begin{equation}
    \sigma^2_i = \frac{1}{n}\sum_{j=1}^{n}X_{i,j}^2 - (\frac{1}{n}\sum_{j=1}^{n} X_{i,j})^2\\
\end{equation}

This reduces the full LayerNorm calculation to two passes; The first pass accumulates a sum and squared sum of each row $\mathrm{i}$. After the row sums are calculated, the $\mathrm{Row Mean_{i}}$ and $\mathrm{1/\sqrt{Row Var_{i}}}$ are calculated using fixed-point arithmetic functions. These constants are fed into the LayerNorm Module shown in Figure~\ref{fig:layernorm}. By passing each row element $\mathrm{j}$ through the LayerNorm Module in a pipelined manner, the final layer-normalized value is efficiently computed.

\subsubsection{Pseudo-Softmax Module}
\label{sec:softmax}

We implement the Pseudo-Softmax function, a hardware-friendly alternative to the Softmax function  proposed in \cite{cardarilli2021pseudo}. We utilize a two-pass approach for computation which is parallelized with matrix multiplication. The Pseudo-Softmax uses base 2 instead of $\mathrm{e}$ to leverage floating point number properties to evaluate exponentiation. 
\begin{equation}
    \widetilde{p}_i = \frac{2^{x_i}}{\sum_{k=1}^{N} 2^{x_k}}
\end{equation}
Let ${a_i}$ be a floating-point number with exponent $x_i$. This removes the need to calculate exponentiation in hardware since it is implicitly handled by the float representation.
\begin{equation}
    \widetilde{p}_i = \frac{a_i}{\sum_{k=1}^{N} a_k}
\end{equation}
The summation term can be expressed as a single floating point number. In this representation, $\text {exp}_{\text {sum}}$ and $\text {mant}_{\text {sum}}$ denote the exponent and mantissa of the resulting number.
\begin{equation} \sum _{k=1}^{N} {a_k}=2^{\text {exp}_{\text {sum}}} \cdot \text {mant}_{\text {sum}}
\end{equation}
The Pseudo-Softmax $\widetilde{p}_i$ for element $x_i$, is calculated as:
\begin{equation}
\widetilde{p}_i=2^{x_i-\text {exp}_{\text {sum}}} \cdot \frac{1}{ 1 \cdot \text {mant}_{\text {sum}}}. 
\label{eq:pseudo-softmax}
\end{equation}

This requires two passes over an input vector $x$. The first pass is to calculate the values $\mathrm{\text {exp}_{\text {sum}}}$ and $\mathrm{\text {mant}_{\text {sum}}}$ and the second pass is to calculate $\widetilde{p}_i$. The reciprocal of the sum mantissa is determined after the floating point sum is computed. Since the Softmax value ranges from 0 to 1, the final result will be the reciprocal bit-shifted by $\text {exp}_{\text {sum}}-x_i + 1$ as per Equation~\ref{eq:pseudo-softmax}. 127 is added to the score row to convert to a floating point exponent which is stored unsigned. A 1 is prepended to the mantissa since it is implicitly present in the floating point format. The result is stored in fixed-point format with only fractional bits to maximize accuracy. The upper 8 bits after the bit-shift contain the fixed-point representation of the Pseudo-Softmax function.

\begin{figure}[t]
    \centering
    \includegraphics[width=0.9\linewidth]{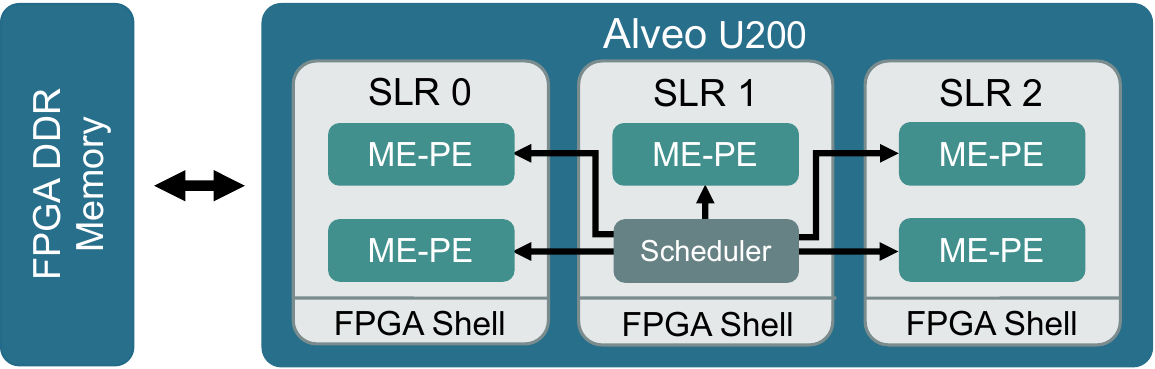}
    \caption{Multi-PE ME-ViT architecture on the Alveo U200.}
    \label{fig:multi-pe}
\end{figure}

\subsection{Multiple ME-PE Architecture for ME-ViT}
\label{sec:multi-pe}

A Multiple ME-PE architecture (Multi-PE) is proposed that contains parallel instantiations of the ME-PE along with a scheduler to coordinate data traffic between them. The Alveo U200 contains 3 Super Logic Regions (SLRs), with SLRs 0 and 2 having 2275 DSPs each and SLR 1 only having 1317. With a $\mathrm{P_{SYS} = 32}$, up to 5 PEs can fit as shown in Figure~\ref{fig:multi-pe}. A smaller $\mathrm{P_{SYS}}$ can fit more SAs in the FPGA, however the BRAM needed for such a design is the same as $\mathrm{P_{SYS} = 32}$ since buffering requirements are unchanged. Therefore, only the $\mathrm{P_{SYS} = 32}$ design is implemented to maximally utilize the available DSPs. Resource utilization is discussed in more detail in Section~\ref{sec:hardware-results}. The Multi-PE architecture can achieve remarkably higher throughput, but an increase in data traffic causes a bottleneck due to the limited 77 GB/s bandwidth to the FPGA DRAM. These results are discussed in Section~\ref{sec:multi-pe-results}.

\section{Evaluation}
\label{sec:eval}
\subsection{Experimental Setup}

\newcolumntype{P}[1]{>{\centering\arraybackslash}p{#1}}

We implement ME-ViT on the Xilinx Alveo U200 platform, consisting of 5867 DSPs, 1766 36k BRAMs, 892K LUTs, and 1831K FFs. It has 4 channels of DDR memory, and a total bandwidth of 77 GB/s. Experimental results are evaluated independently for each ME-ViT mode, and theoretical values are presented which remove extra latencies added from Vitis synthesis and Place and Route (P\&R). All implementations are designed for 300 MHz, and 150 MHz figures are provided to compare with other designs. The ME-PE is designed and evaluated using Vitis HLS 2023.1. Power estimates are calculated using AMD Power Design Manager 2023.1.1. 

A single ME-PE is analyzed on four common ViT model sizes shown in Table~\ref{table:model-variants}. ME-PEs with systolic array size $\mathrm{P_{SYS}=32}$ and $\mathrm{P_{SYS}=16}$ are analyzed to provide insight into the performance scalability across FPGAs of varying DSP resources. As the size of the systolic array decreases, there is a corresponding reduction in total FPS (frames per second). Memory bandwidth also reduces despite smaller systolic arrays requiring more frequent data transfers. This relationship between scale and memory bandwidth is explored in Section~\ref{sec:mbw-comparison}. Finally, Multi-PE results are calculated based on single $\mathrm{P_{SYS} = 32}$ ME-PE performance to maximally utilize the available DSPs.

\begin{table}[h]
\centering
\caption{Model Variants}
\begin{tabular}{cP{1cm}P{1.3cm}P{1cm}cP{1.25cm}}
\toprule
\textbf{Model} & \textbf{Image Size} & \textbf{Model Dimension} & \textbf{Num Heads} & \textbf{Layers} & \textbf{Parameter Count} \\
\midrule
\textbf{ViT-B} & 256\textsuperscript{2} & 768 & 12 & 12 & 86M \\
\textbf{DeiT-B} & 224\textsuperscript{2} & 768 & 12 & 12 & 86M \\
\textbf{DeiT-S} & 224\textsuperscript{2} & 384 & 6 & 12 & 22M \\
\textbf{DeiT-T} & 224\textsuperscript{2} & 192 & 3 & 12 & 6M \\
\bottomrule
\end{tabular}
\label{table:model-variants}
\end{table}

ViT variants shown in Table~\ref{table:model-variants} are evaluated on the ME-ViT architecture. ViT-B refers to the base ViT model presented in \cite{dosovitskiy2021image} but with 256 input image resolution. DeiT-B, DeiT-S, and DeiT-T refer to model sizes presented in \cite{touvron2021training}, all on 224 input image resolution. The key difference between DeiT models lies in their respective model dimensions:
768, 384, and 192.

\subsection{Results on Hardware}
\label{sec:hardware-results}

Results for hardware utilization are shown in Table~\ref{tab:resource-util}. The three ME-ViT modes are implemented separately, and throughput values in Table~\ref{tab:performance} are derived from the latency per mode. Since each mode largely utilizes the same resources but with different control logic, the unified design's resources would marginally exceed that of the largest mode (MLP Mode). Resources are shown for the $\mathrm{P_{SYS} = 32}$ ME-PE for the ViT-B model. Resource consumption is unchanged for DeiT-B, and BRAM usage drops to 176 and 144 for DeiT-S and DeiT-T respectively. For $\mathrm{P_{SYS} = 16}$, 256 DSPs are used, with other values remaining the same as buffering requirements do not change.

\begin{table}[h]
\centering
\caption{Hardware Resource Utilization}
\begin{tabular}{ccccc}
\toprule
\textbf{Hardware Configuration} & \textbf{DSP} & \textbf{BRAM36} & \textbf{LUT (K)} & \textbf{FF (K)} \\
\midrule
\textbf{LP Mode} & 1024 & 288 & 159 & 93 \\
\textbf{MSA Mode} & 1024 & 288 & 166 & 107 \\
\textbf{MLP Mode} & 1024 & 288 & 192 & 132 \\
\textbf{Auto Vit Acc \cite{li2022autovitacc}} & 2066 & -- & 128 & -- \\
\bottomrule
\end{tabular}
\label{tab:resource-util}
\end{table}

\newcolumntype{Y}{>{\centering\arraybackslash}p{1.2cm}} 
\newcolumntype{Z}{>{\centering\arraybackslash}p{0.15\textwidth}} 
\setlength{\tabcolsep}{4pt}
\begin{table*}[t]
\centering
\caption{Platform Performance Comparison}
\label{tab:performance}
\begin{tabularx}{\textwidth}{cYYYYYYYYYYY}
\toprule
\multirow{2}{*}{\textbf{Platform}} & 
\multirow{2}{*}{\parbox{1cm}{\centering \textbf{CPU i7-9800X \cite{sun2022vaqf}}}} & 
\multirow{2}{*}{\parbox{1.2cm}{\centering \textbf{GPU Titan RTX \cite{sun2022vaqf}}}} & 
\multirow{2}{*}{\parbox{1.2cm}{\centering \textbf{GPU Jetson TX2 \cite{li2022autovitacc}}}} & 
\multirow{2}{*}{\parbox{1.2cm}{\centering \textbf{Auto Vit Acc \cite{li2022autovitacc}}}} & 
\multirow{2}{*}{\parbox{1cm}{\centering \textbf{ViTA \cite{nag2023vita}}}} &
\multicolumn{2}{c}{\parbox{2.4cm}{\centering\textbf{ME-ViT}}} & 
\multicolumn{2}{c}{\parbox{2.4cm}{\centering\textbf{ME-ViT Theoretical}}} & 
\multicolumn{2}{c}{\parbox{2.4cm}{\centering\textbf{Multi-PE}}} \\
\cmidrule(lr){7-8} \cmidrule(lr){9-10} \cmidrule(lr){11-12}
& & & & & & \textbf{150 MHz} & \textbf{300 MHz} & \textbf{150 MHz} & \textbf{300 MHz} & \textbf{150 MHz} & \textbf{300 MHz} \\
\midrule
\textbf{Latency (ms)} & 65.35 & 5.45 & 127 & 38.61 & 363.64 & 83.38 & 41.69 & 75.73 & 37.86 & 75.73 & 37.86 \\ 
\textbf{FPS} & 15.3 & 183.4 & 7.87 & 25.9 & 2.75 & 11.99 & 23.98 & 13.20 & 26.40 & 66.02 & 132.04 \\ 
\textbf{Power (W)} & 100 & 260 & 12.28 & 9.4 & 0.88 & 6.5 & 9.3 & 6.5 & 9.3 & 17.8 & 31.8 \\ 
\textbf{FPS/Watt} & 0.15 & 0.71 & 0.64 & 2.76 & 3.13 & 1.84 & 2.57 & 2.03 & 2.83 & 3.71 & 4.15 \\
\textbf{FPS/DSP} & -- & -- & -- & 0.012 & -- & 0.012 & 0.023 & 0.013 & 0.026 & 0.013 & 0.026 \\
\bottomrule
\end{tabularx}
\end{table*}

\subsection{Performance Comparison}

Performance comparison values for ME-ViT are shown in Table~\ref{tab:performance}. Theoretical values are presented which are calculated by leveraging extra parallelism which cannot be achieved through Vitis HLS. These figures are achievable with Verilog synthesis. All synthesized designs achieve an operating frequency of 300 MHz, however, competing solutions (Auto Vit Acc~\cite{li2022autovitacc} and ViTA~\cite{nag2023vita}) are implemented at lower frequencies. To accurately compare design metrics, 150 MHz figures are provided.

We compare ME-ViT performance against four baseline platforms: (1) CPU only platform (Intel i7-9800X), (2) high-power GPU accelerated platform (Nvidia Titan RTX), (3) low-power GPU accelerated platform (Nvidia Jetson TX2), (4) FPGA ViT accelerator presented in Auto Vit Acc \cite{li2022autovitacc}, and (5) Edge FPGA ViT accelerator presented in ViTA~\cite{nag2023vita}.

A single ME-PE achieves a theoretical throughput of 26.4 FPS, a 1.04$\mathrm{\times}$ improvement over the Auto Vit Acc implementation. Notably, Auto Vit Acc is implemented at 150 MHz which outperforms a single ME-PE at the same clock frequency due to higher DSP usage. The ME-PE has a similar FPS/DSP efficiency as Auto ViT Acc at 150 MHz, and sees a 2.16$\mathrm{\times}$ improvement at 300 MHz. The ME-PE has a high power efficiency of 2.83 FPS/Watt, outperforming all other platforms except for ViTA. In comparison to a GPU with a similar power usage (TX2), ME-ViT has a 4.42$\mathrm{\times}$ improvement in FPS/Watt, demonstrating the high efficiency of the custom architecture approach.

\begin{figure}[t!]
    \centering
    \includegraphics[width=\linewidth]{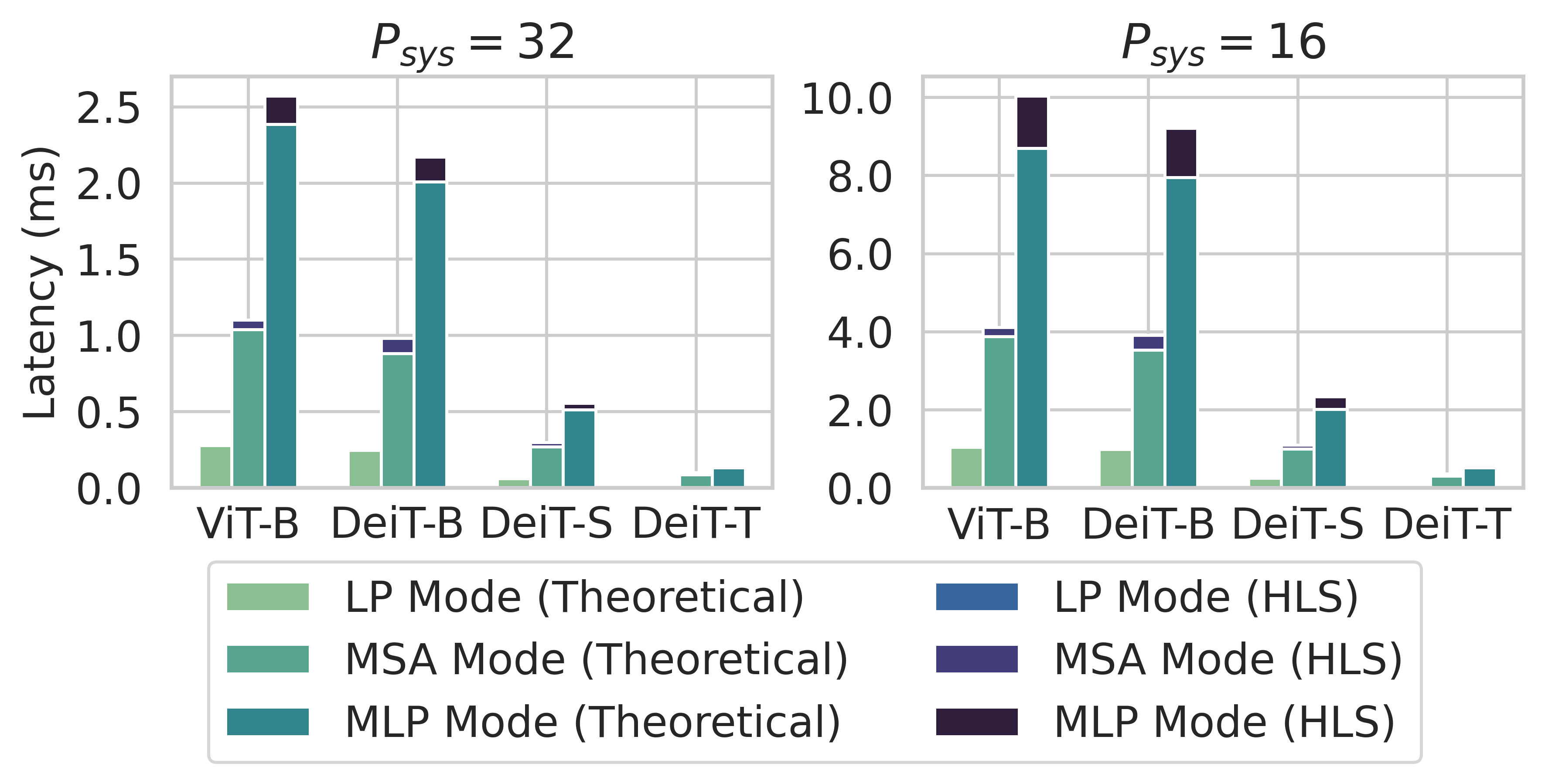}
    \caption{Breakdown of latency per mode for ME-ViT on various models. Vitis implementation (HLS) introduces small extra latency overhead to the theoretical performance.}
    \label{fig:br_latency}
\end{figure}

\subsection{Overall Throughput and Latency}

Overall throughput measured in frames per second (FPS) for the 4 models is shown in Table~\ref{tab:model-performance}. FPS improves as model sizes get smaller, and $\mathrm{P_{SYS} = 16}$ experiences roughly 0.25$\times$ the throughput of corresponding $\mathrm{P_{SYS}=32}$ designs. Latencies per ME-ViT mode are presented in Figure~\ref{fig:br_latency}. MLP Mode exhibits the longest duration out of the three modes, taking approximately 60 percent of execution time across all models. A reduction of input image size from 256 to 224 results in an average 1.17$\times$ improvement for $\mathrm{P_{SYS} = 32}$ and an average 1.08$\times$ improvement for $\mathrm{P_{SYS} = 16}$. For both systolic array sizes, a reduction of model dimension in half results in an average improvement of 3.7$\times$.  

\begin{figure}[t!]
    \centering
    \includegraphics[width=1\linewidth]{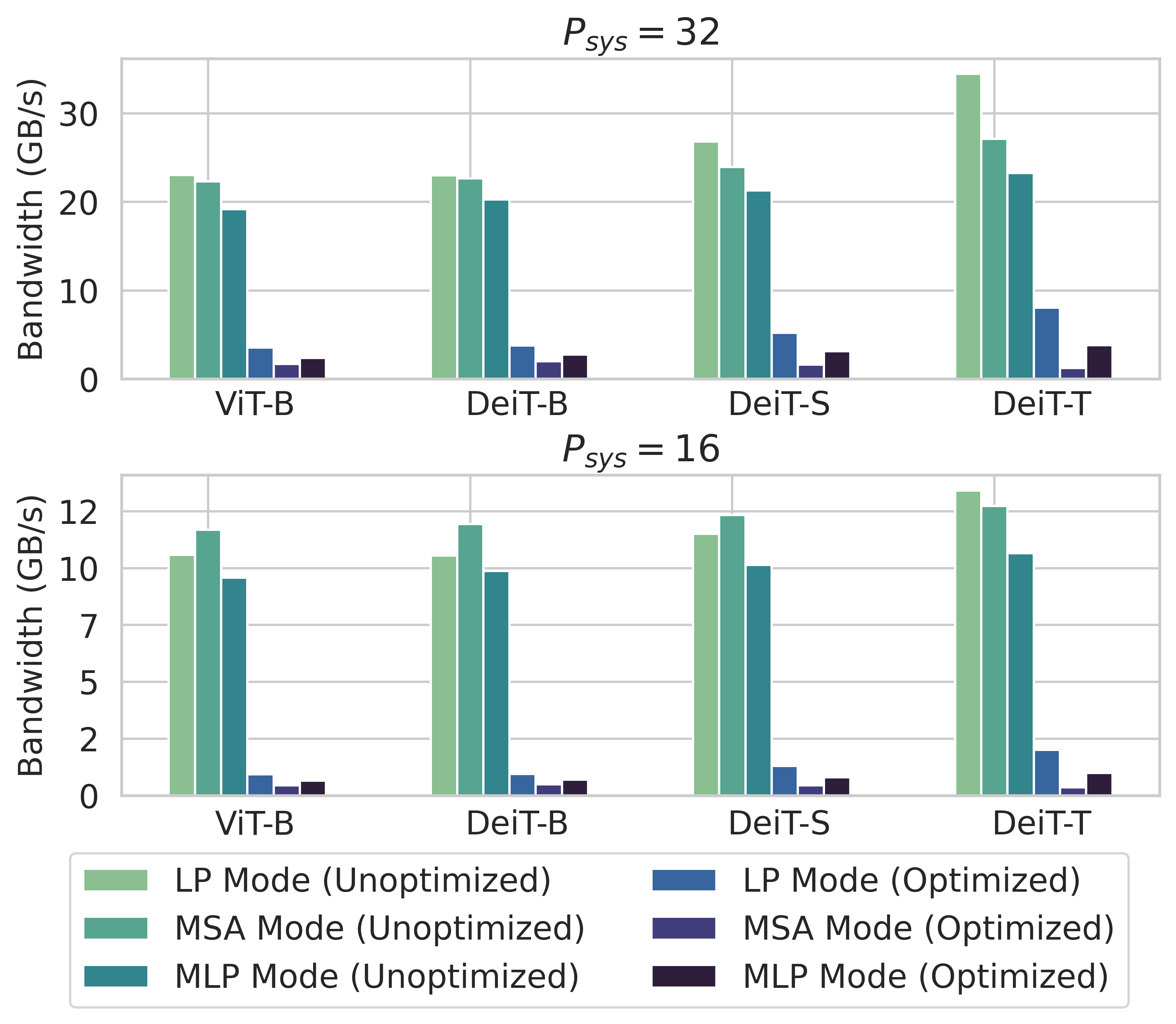}
    \caption{Breakdown of bandwidth per mode for ME-ViT on various models, lower is better.}
    \label{fig:br_bandwidth}
\end{figure}

\begin{table}[h]
\centering
\caption{Model Performance (FPS)}
\begin{tabular}{cP{1.4cm}P{1.4cm}P{1.4cm}P{1.4cm}}
\toprule
\textbf{Model} & \textbf{HLS ($\mathrm{P_{SYS}}$=$\mathrm{32}$)} & \textbf{Theoretical ($\mathrm{P_{SYS}}$=$\mathrm{32}$)} & \textbf{HLS ($\mathrm{P_{SYS}}$=$\mathrm{16}$)} & \textbf{Theoretical ($\mathrm{P_{SYS}}$=$\mathrm{16}$)}\\
\midrule
\textbf{ViT-B} & 20.64 & 22.38 & 5.40 & 6.08 \\
\textbf{DeiT-B} & 23.98 & 26.40 & 5.81 & 6.64 \\ 
\textbf{DeiT-S} & 87.64 & 98.25 & 22.13 & 25.53 \\ 
\textbf{DeiT-T} & 298.52 & 352.27 & 78.55& 94.13 \\
\bottomrule
\end{tabular}
\label{tab:model-performance}
\end{table}

\subsection{Memory Bandwidth Comparison}
\label{sec:mbw-comparison}

As there exist no published results on memory bandwidth for similar FPGA architectures, a non optimized approach is calculated with the following characteristics that are common in various designs \cite{sun2022vaqf, hu2021vistop, nag2023vita}: Each BMM loads two input matrices. If an input block matrix was used for the previous multiply, it remains loaded. 2) All calculated matrix blocks are written back to DRAM. 3) Softmax and LayerNorm\cite{ba2016layer} are calculated on the CPU and are implicitly included in intermediate write-backs.

\begin{table}[t!]
\centering
\caption{Memory Bandwidth Improvement}
\begin{tabular}{cP{1.4cm}P{1.4cm}P{1.4cm}P{1.4cm}}
\toprule
\textbf{Model} & \textbf{Total ($\mathrm{P_{SYS}}$=$32$)} & \textbf{Peak ($\mathrm{P_{SYS}}$=$\mathrm{32}$)} & \textbf{Total  ($\mathrm{P_{SYS}}$=$\mathrm{16}$)} & \textbf{Peak  ($\mathrm{P_{SYS}}$=$\mathrm{16}$)}\\
\midrule
\textbf{ViT-B} & 9.22 & 13.07 & 17.14 & 25.58 \\
\textbf{DeiT-B} & 8.25 & 11.29 & 16.62 & 23.79 \\ 
\textbf{DeiT-S} & 7.06 & 14.60 & 17.53 & 27.60 \\ 
\textbf{DeiT-T} & 8.77 & 21.28 & 17.89 & 35.29 \\
\bottomrule
\end{tabular}
\label{tab:memory-bandwidth}
\end{table}

Figure~\ref{fig:br_bandwidth} shows memory bandwidth figures for ME-ViT on all four models for both $\mathrm{P_{SYS} = 32}$ and $\mathrm{P_{SYS} = 16}$. Total and peak bandwidth improvements are shown in Table~\ref{tab:memory-bandwidth}. Peak improvement occurs in the MSA Mode as this has the most back-and-forth traffic in the unoptimized case. Despite fewer model parameters to transfer, a higher improvement is seen on smaller models since less time is spent on computation and therefore a larger proportion of data movement needs to occur in the same time. There is an approximate 3.8$\times$ reduction in latency between $\mathrm{P_{SYS} = 32}$ and $\mathrm{P_{SYS} = 16}$, but an approximate 2$\times$ reduction in data transferred. This results in a larger reduction in memory bandwidth for $\mathrm{P_{SYS} = 16}$.

\subsection{Systolic Array Size Comparison}
\label{sec:sys-array-res}

The systolic array has a large impact on the computational efficiency, defined as the total computation performed divided by the minimum computation required. Depending on how $\mathrm{P_{SYS}}$ divides both the model dimension and the layer height, computational efficiency can greatly vary. When $\mathrm{P_{SYS}}$ poorly matches these dimensions, the BMMs at the right and bottom boundaries fill only a small portion of the systolic array, leading to wasted computation.

\begin{figure}[t!]
    \centering
    \includegraphics[width=\linewidth]{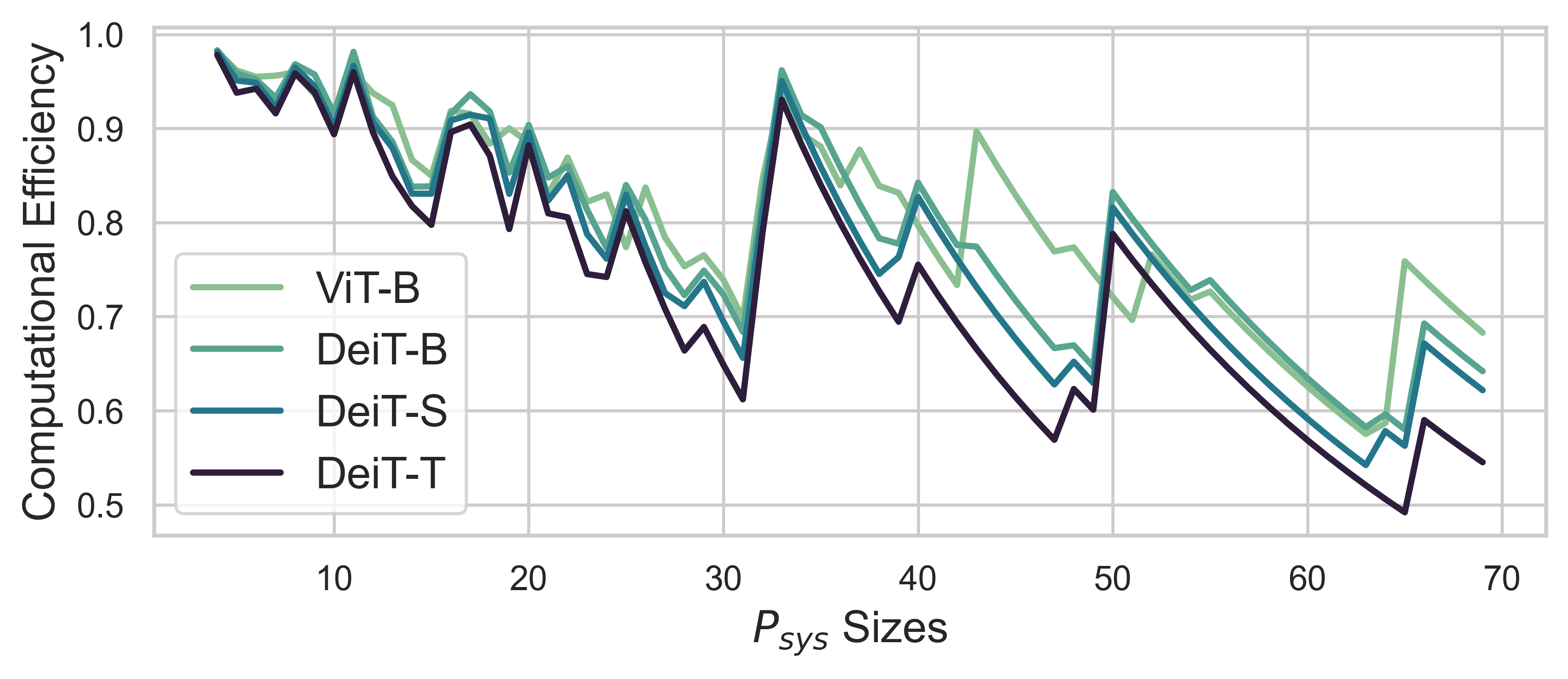}
    \caption{Computational Efficiency vs. $P_{SYS}$ for all models.}
    \label{fig:comp-eff}
\end{figure}

As shown in Figure~\ref{fig:comp-eff}, there are periodic cycles in efficiency ranging from 0.95 to 0.5, with notable peaks occurring at 11, 17, 33, 50, and 66. In addition, simply increasing  $\mathrm{P_{SYS}}$ leads to an overall downward trend in efficiency. These results underscore the importance of tuning hardware to fit the model dimensions.

\subsection{Multi-PE ME-ViT Performance}
\label{sec:multi-pe-results}

\begin{figure}[t!]
    \centering
    \includegraphics[width=\linewidth]{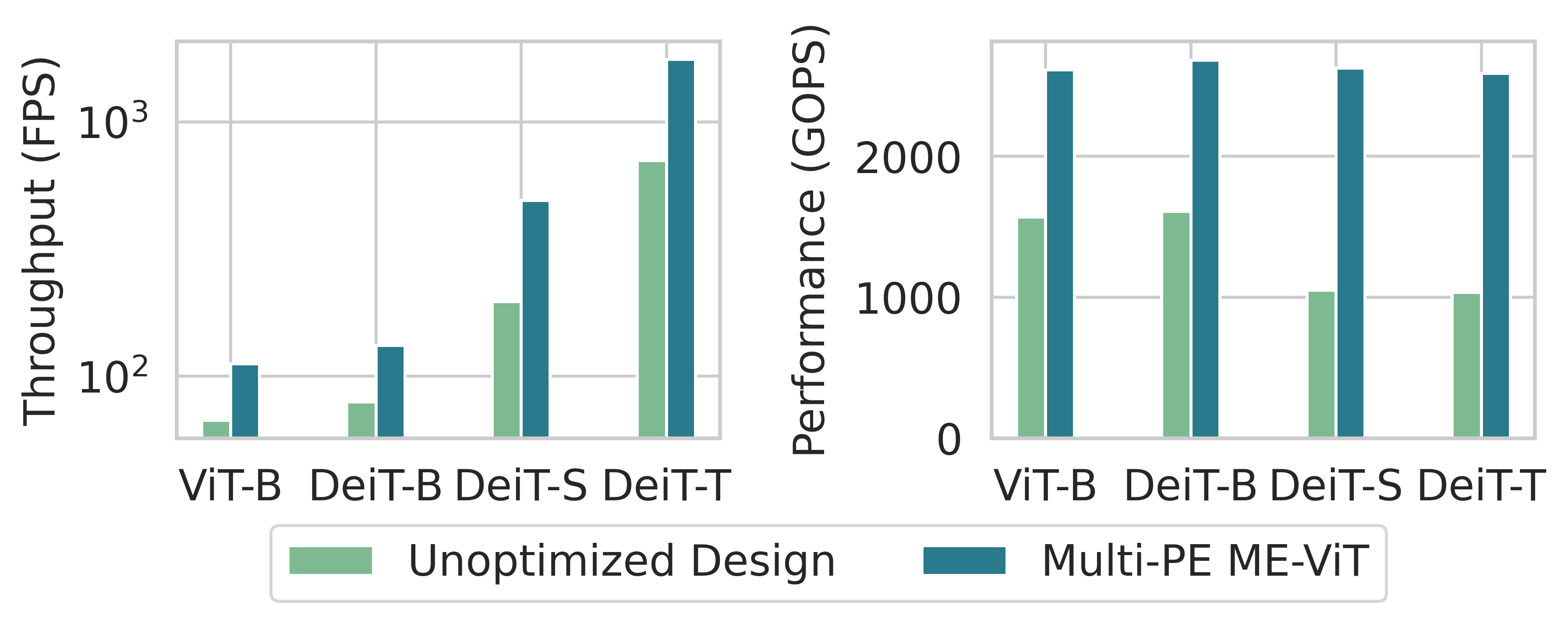}
    \caption{Throughput (left) and performance (right) comparison between Multi-PE and the unoptimized design, higher is better. Throughput vertical axis is in log scale.}
    \label{fig:multi-pe-perf}
\end{figure}

We analyze the performance of the Multi-PE to determine the effectiveness of ME-ViT when scaled to larger FPGAs. Multi-PE performance results are shown in Figure~\ref{fig:multi-pe-perf}. The unoptimized design for ViT-B and DeiT-B can only support 3 ME-PEs before performance is limited by memory bandwidth. For DeiT-S and DeiT-T, the unoptimized design can only support 2 ME-PEs. ME-ViT allows 5 ME-PEs to be supported for all models, resulting in a 1.66$\times$ improvement in both FPS and GOPS (Giga Operations per Second) for ViT-B and DeiT-B, and a 2.5$\times$ improvement for DeiT-S and DeiT-T. The theoretical maximum GOPS for 5 ME-PEs is 3072, yet a maximum of 2682 is achieved due to inefficiencies with irregularly-sized matrix multiplication. Matrix blocks that do not completely fill the systolic array result in unused DSPs, leading to an overall reduction in GOPS.

For larger ViT models such as ViT-Large and ViT-Huge ($\mathrm{D=1024}$ and $\mathrm{1280}$), total BRAM usage will increase to 384 and 608 respectively with $\mathrm{P_{sys} = 32}$. ME-PEs of this size can still fit inside a large FPGA like the Alveo U200, but fewer total will fit in the Multi-PE design. This results in a large number of unused DSPs, reducing the total GOPs from the theoretical maximum.

\section{Conclusion}
\label{sec:conclusion}
In this paper, we proposed ME-ViT, a novel ViT hardware accelerator that mitigates the high-bandwidth needs of ViT inference. ME-ViT minimizes the memory traffic for a ViT accelerator on an FPGA through a single-load policy and multi-purpose buffers within a memory-efficient processing element (ME-PE). ME-ViT achieves up to a 17.89$\times$ overall improvement in memory bandwidth, and up to a 2.16$\times$ improvement in throughput per DSP over state-of-the-art ViT accelerators on FPGA.  ME-ViT enables implementation of up to 5 ME-PEs on a Xilinx Alveo U200, achieving a 5.10$\times$ improvement in throughput over the FPGA baseline. Future research will focus on extending the ideas in this paper beyond ViTs to Large Language Models which inhibit a single-load policy due to limited on-chip memory.


\section*{Acknowledgment}

This work has been supported by the U.S. National Science Foundation under grant numbers CNS-2009057 and SaTC-2104264.
We are grateful to Bingyi Zhang for his insightful perspectives in the development of this paper.

\bibliographystyle{./bibliography/IEEEtran}
\bibliography{./bibliography/reference.bib}
\end{document}